\documentclass{article}[10pt]

\usepackage{bbm,times,amsmath}

\begin{document}

\bigskip\bigskip\bigskip

\noindent {\large{\bf Nontrivial quantum effects in biology: 
A skeptical physicists' view}}
\bigskip\bigskip

\noindent {{H.M.\ Wiseman$^{1}$ and J.\ Eisert$^{2,3}$}

\smallskip\smallskip

\noindent
{\footnotesize{\it 
\renewcommand{\baselinestretch}{0.75}
$^1$ Centre for Quantum Computer Technology\\
Griffith University\\
Brisbane 4111, Australia\smallskip\\
$^2$ Blackett Laboratory\\
Imperial College London\\
London SW7 2BW, UK\smallskip\\
$^3$ Institute for Mathematical Sciences\\
Imperial College London\\
Exhibition Road\\
London SW7 2BW, UK }}\\

\newcommand{\R}{\mathbbm{R}}
\newcommand{\rr}{\mathbbm{R}}
\newcommand{\nn}{\mathbbm{N}}
\newcommand{\cc}{\mathbbm{C}}
\newcommand{\id}{\mathbbm{1}}

\newcommand{\tr}{{\rm tr}\,}
\newcommand{\gr}[1]{\boldsymbol{#1}}
\newcommand{\be}{\begin{equation}}
\newcommand{\ee}{\end{equation}}
\newcommand{\bea}{\begin{eqnarray}}
\newcommand{\eea}{\end{eqnarray}}
\newcommand{\ket}[1]{|#1\rangle}
\newcommand{\bra}[1]{\langle#1|}
\newcommand{\avr}[1]{\langle#1\rangle}
\newcommand{\G}{{\cal G}}
\newcommand{\eq}[1]{Eq.~(\ref{#1})}
\newcommand{\ineq}[1]{Ineq.~(\ref{#1})}
\newcommand{\sirsection}[1]{\section{\large \sf \textbf{#1}}}
\newcommand{\sirsubsection}[1]{\subsection{\normalsize \sf \textbf{#1}}}

\newcommand{\proofend}{\hfill\fbox\\\medskip }

\medskip

\newcommand{\sch}{Schr\"odinger}
\newcommand{\ea}{{\em et al.}}

\begin{quotation}
When you have excluded the trivial, whatever remains, however improbable, 
must be a good topic for a debate.\footnote{Apologies to A.C.\ Doyle.}
  \end{quotation}

\section{Introduction}

This chapter is somewhat of an anomaly in this book. Firstly, its authors profess no particular knowledge of any effects in biology, (whether quantum or non-quantum, trivial or non-trivial), both being theoretical quantum physicists by trade. Secondly, we adopt here a skeptical view on the existence of such effects if they fall in the non-trivial class. That two such skeptical non-experts have been invited to contribute to this volume came about as a result of the public debate (reproduced earlier in this volume) at the 2nd International Symposium on Fluctuations and Noise, held in the Canary Islands in 2004. We two were invited by Derek Abbott to affirm the statement that {\em Quantum effects in biology are trivial.} 

This chapter will reproduce many of the arguments that we put in that debate, although hopefully somewhat more coherently than we communicated them at the time. It also contains some arguments that were not covered in the debate.  Obviously the debate would have been pointless unless both sides had agreed on what counts as a {\em non-trivial} quantum effect in biology. Thankfully, all participants in the debate did agree, more or less, although only one (HMW) offered a formal definition: that a non-trivial quantum effect in biology is one that would convince a biologist that they needed to take an advanced quantum mechanics course and learn about Hilbert space \index{Hilbert space} and operators etc., so that they could understand the effect. 

To use the word  ``trivial'' to characterize all quantum effects in biology that do not increase enrollments of biologists in advanced quantum physics courses is unfortunate. Neither we, nor, we imagine, any of the debate participants, wish to denigrate the interesting and challenging research into quantum effects relevant to biology such as coherent excitations \index{coherent excitations} of biomolecules \cite{Hel02,GilMcK05}, quantum tunneling \index{quantum tunneling} of protons \cite{KohKli99}, van der Waals forces \index{van der Waals forces} \cite{Par05}, ultrafast dynamics through conical intersections \index{conical intersections} \cite{CedGinBur05}, and phonon-assisted electron tunneling as the basis for our sense of smell \index{smell, sense of} \cite{Nose07}. 
But here we are concerned not with these real (or at least plausible) 
quantum effects, but rather with more exotic, unproven (and, we believe, implausible) effects.

What might these non-trivial sorts of quantum effects be? Several have been suggested in the literature (see other papers in this volume), but we will concentrate upon four: 
%\begin{itemize}
%\item 
A quantum life principle; 
%\item 
Quantum computing in the brain; 
%\item 
Quantum computing in genetics; and 
%\item 
Quantum consciousness.
%\end{itemize}
These intriguiging topics provide the structure of our chapter. We devote one section each to 
briefly explaining, and then arguing the implausibility of, these hypothetical effects. 
It is hence the purpose of the present chapter to be cautionary: 
 to warn of ideas that are more appealing at first sight than they are realistic. 

We end, however, on a more constructive note in our final section, by pointing 
out that there is one sense in which it seems likely that quantum effects introduce a non-trivial difference between brains and digital computers. 
This section (Quantum free will) is of interest philosophically \index{philosophy} rather than 
scientifically, so we do not see at as an exception to our claim that 
biologists should not want to enroll in advanced quantum physics 
courses.\footnote{Philosophers, on the other hand, should!}

\section{A quantum life principle} \index{life principle}

\subsection{A quantum chemistry principle?}

It is a widely held belief that the origin of life \index{origin of life} is extremely unlikely according to established science. This has led some to argue that there exists a natural principle, in addition to existing laws, that guarantees that life must arise in the Universe --- see Ref.~\cite{Dav04a}. \index{Davies, Paul}
In this review, Davies points out difficulties with this argument, but apparently he gives it 
some credibility since he used it in the 2004 Canary Island debate. There he stated that, unless life is miraculous, there must be a life principle, and that 
since it is a fundamental physical principle, it must be related to our most fundamental physical theory: quantum mechanics. In Ref.~\cite{Dav04a}  he suggests that the origin of life may be related to quantum search algorithms, \index{search algorithms} an idea we discuss in Sec.~\ref{sec:WEgen}.

That a belief is widely held does not make it correct. Indeed, we claim that the origin of life is entirely plausible according to established physical theories. Moreover, the relevant physical theory, chemistry, \index{chemistry} has no deep relation to quantum physics. To understand chemical structure and reactions at a fundamental level it is, of course, necessary to use quantum physics. But chemistry is usually regarded as emerging from physics in a straight-forward (upwardly causal, not downwardly causal \cite{Dav04b}) way. If this were not the case, it would be necessary to postulate not merely a ``quantum life principle'', but also a ``quantum chemistry principle'' (along with, presumably, a ``quantum condensed matter principle'', a ``quantum atom principle'', and so on).

That life is an epiphenomenon of chemistry, and one whose appearance on earth is unsurprising, even expected, is well argued by Dawkins in his most recent popular book 
on evolution \cite{Daw04}. First, he stressses (pp.~575-81) that the essence of life, the aspect of life that must precede all others, is {\em heredity}. \index{heredity} Heredity means the existence of a varied population of replicators in which the variation is (at least partially) inherited in the act of replication. To quote Dawkins, \index{Dawkins, Richard}
\begin{quotation}
%Theories of the origin of life need to account for both heredity and metabolism, but some writers have mistaken the priority. They 
[S]ome writers \ldots have sought a theory of metabolism's \index{metabolism} spontaneous origin, and somehow hoped that heredity would follow, like other useful devices. But heredity \ldots is not to be thought of as a useful device. Heredity has to be on the scene first because, before heredity, usefulness itself has no meaning. Without heredity, and hence natural selection, 
\index{natural selection}  there would have been nothing to be useful for. 
%The very idea of usefulness cannot begin until the natural selection of hereditary information does.
\end{quotation}

Accepting Dawkin's imperative, the origin of life can be illuminated by seeking the simplest carrier of hereditary information in the natural world. A well publicized example \cite{Dav99,Daw04} is {\em Spiegelman's Monster}, \index{Spiegelman's Monster} named after its developer \cite{Spi72}. It is far simpler than the viruses from which it was derived. It is an RNA \index{RNA} strand a mere 218 nucleotides long; that is, it is a large molecule containing less than $10^4$ atoms. The environment in which it replicates is an aqueous solution of activated nucleotides, plus an enzyme Q$\beta$-replicase. As shown by Spiegelman and Orgel, the monster carries hereditary information, and undergoes natural selection \footnote{Indeed, later work \cite{Eig97} showed that, through natural selection, the monster reduced even further in size, down to a mere 50 nucleotides --- a few thousand atoms!}. Most remarkably, self-replicating monsters appear {\em spontaneously} in the environment described above \cite{SumLuc75}.

The point of these investigations is not that Spiegelman's monster is the first life --- that probably developed in quite different environments \cite{Dav99,Daw04}. Rather, in the present context, the points are: (i) that the beginnings of life need not be nearly so complicated as is imagined by those who stress its implausibility; and (ii) that nothing in these experiments suggest that anything other than chemistry is involved in the self-assembly and replication of these largish molecules. Indeed, it is likely that the chemical reactions involved could be reproduced, in the not too distant future, by simulations based at the atomic level. Such a simulation would be a definitive refutation of the idea of a quantum life principle. 

\subsection{The anthropic principle}\index{anthropic principle}

It could be argued that, even if life is an almost inevitable consequence of chemistry in suitable environment, this fact itself requires explanation. That is, does it not seem miraculous that the physical world enables life to arise? Specifically, it has been argued that the fundamental constants of physics are ``fine-tuned'' so as to allow the existence of long-lasting stars, planets, liquid water {\em etc.} that are apparently necessary for life to arise \cite{BarTip86}. Such an argument is known as the strong anthropic principle \footnote{Martin Gardiner has suggested the name Completely Ridiculous Anthropic Principle for the more extreme versions of this  principle  \cite{Gar86}.}. According to the standard model \index{standard model} of particle physics, there are some 20 fundamental constants whose values are arbitrary, and according to 
%speculative 
theories like string theory, \index{string theory} these ``constants'' are in fact quantum variables \cite{Sus05}. Thus it might seem plausible to claim that life is somehow linked to quantum cosmology \cite{Dav04a}.

Leaving aside the possible lack of imagination of physicists with regard to the sorts of universes in which life may exist, it seems unnecessary to invoke the strong anthropic principle to argue for a quantum life principle, when the weak anthropic principle has just as much explanatory power. The weak anthropic principle simply states that we should condition all our predictions on the undeniable fact that we are here to ask the question \cite{BarTip86}. Thus, if asked, what is the chance that the fundamental constants will be found to have values that enable life to evolve, we would have to say that the chance is essentially unity, since life evidently has evolved. That is, invoking some special principle to explain that life must have appeared and evolved intelligent observers is as unnecessary as invoking a special principle to explain that the language we call English must have developed and become widespread.

\section{Quantum computing in the brain}

\subsection{Nature did everything first?}

 In the past decade of so,  the field of 
 quantum information \index{quantum information} (theory and experiment) has exploded \cite{NieChu00}. 
 This is driven largely by the prospect of building a large-scale quantum computer, \index{quantum computing}
 that could compute much faster than any conceivable classical computer by existing in a superposition of different computational states.\index{superposition}
 This leads naturally to the conjecture that the brain itself may be a quantum computer \cite{Hagan}. 
  
When looking at the wealth of existing
life forms, the following observation becomes 
apparent: Nature had the idea first. Indeed, in nature we can 
find parachutes and explosives, surfaces
reminiscent of the most sophisticated nanostructured
materials used in aeronautics today 
to reduce the aerodynamic resistance. Many 
effects and concepts of physics can indeed
be found to be exploited by some life form to its benefit.
So, after all, why should this not apply to the
brain being a  quantum computer?

We would argue that this is not a legitimate argument.
While it is striking that some features have been ``invented''
by nature, the argument as such is a ``postselected
argument'', based on case studies of anecdotal 
character. It is equally (if not more) easy to collect counterexamples
of the same character; that is, inventions for which no 
counterpart in nature is known. 
For example, there are no metal skeletons, despite 
metal being much stronger than bone. There is no
radio (long distance) communication, albeit this certainly
being  a useful and feasible means of communication.
No closed-cycle refrigeration  based on gas expansion is known. 
There is no use of inferometry to measure distances. 
Also, the eye is as such a really lousy camera, corrected
by the ``software'' of the brain.  
This last example illustrates 
a general point: nature makes do with things that are good enough; it 
does not do precision engineering. If there is one thing 
a quantum computer requires, it is precision, as we discuss below in Sec.~\ref{Sec:QEC}.

\subsection{Decoherence as make or break issue}

The case for the brain being a quantum computer, or indeed for 
quantum mechanics playing any key role 
at a macroscopic level in the nervous system, is weakest 
because of one effect: decoherence \cite{Dec,Zur98}. \index{decoherence}

A quantum system is never entirely isolated from its environment,
which is always ``monitoring'' its dynamics. That is, information is transferred
into the environment, where it is diluted into a larger and larger 
number of degrees of freedom. As a result, superposition \index{superposition} states become,  
for all practical purposes, indistinguishable from classical mixtures \index{mixture} of 
alternatives on a time scale known as the decoherence time \index{decoherence time} \cite{Dec,Zur98,Eis}. 
In short, quantum coherence is lost, as an effect
of the environment monitoring the system. 

This effect of decoherence is one of the 
main concerns in research on quantum computation \cite{NieChu00}, where
ingenious ways are being explored of shielding engineered and strongly 
cooled quantum systems from their respective environments. 
In fact, decoherence is the key challenge in  the realization of a full-scale quantum computer. In large scale biological systems, like the brain, 
decoherence renders large scale 
coherence (as necessary for quantum computation) very implausible. 
Even the most optimistic 
researchers cannot deny the fact that the brain is a warm and wet environment.
This is in contrast to the high-vacuum environment used in the 
beautiful experiments on spatial superpositions of organic 
molecules from Markus Arndt's and 
Anton Zeilinger's group in Vienna \cite{Zei03}.
\index{Zeilinger, Anton}
In the realistic biological setting, even the most conservative 
upper bounds to realistic decoherence times are 
dauntingly small \cite{Tegmark}. 

It is essential to keep in mind that large-scale 
quantum computation does not mean merely 
computing with a large number of systems, each of which behaves quantum mechanically. 
If  coherence prevails only in 
subsystems of a quantum computer, but not over wide parts 
of the whole system,
the computation would be no more powerful than its classical 
counterpart. Simply putting subsystems together
operating on quantum rules with no coherence between them cannot 
give rise to a quantum computer \cite{NieChu00}. 
To create the large scale superpositions necessary for quantum computation requires
preserving coherence for a long time, \index{quantum coherence} long enough to enable all 
the different subsystems to interact. 

Tegmark's article \cite{Tegmark} \index{Tegmark, Max}
is a careful discussion of the plausibility of 
preserving coherence over long times under the conditions in the brain.
He focuses on two situations where it has been suggested that quantum superpositions
could be maintained:  a superposition of a neuron \index{neuron} firing or not \cite{Schade}; 
and a superposition
of kink-like polarization excitations in microtubules, \index{microtubules} playing a central role in the proposal of 
Hameroff and Penrose \cite{Ham}. \index{Hameroff, Stuart} \index{Penrose, Roger}
The firing of a neuron is a complex dynamical process of a chain reaction, 
involving Na$^+$ and K$^+$ ions to quickly flow across a membrane. Tegmark provides
a conservative estimate of the relevant 
decoherence times for a coherent superposition of a neuron firing including only
the most relevant contributions, arriving at a number of $10^{-20}$ seconds. Similarly,
he discusses decoherence processes in microtubules, hollow cylinders of long 
polymers forming the cytoskeleton of a cell. Again, a conservative estimate 
gives rise to an estimated time of $10^{-13}$ seconds on which superpositions 
decohere to mere mixtures.\footnote{To be fair, it should be noted that 
Hagan \ea\ \cite{Hagan}
themselves argue that decoherence times may be significantly shorter than this \cite{Rosa}.}
The general picture from a discussion of decoherence times
that emerges is the following: Even if superposition 
states were to appear in the
processes relevant for brain functioning, they would persist for times that fall short 
(by many orders of magnitude!) of the time scales 
necessary for the proposed quantum effects to become 
relevant for any thought processes.

\subsection{Quantum error correction} \label{Sec:QEC} \index{quantum error correction}

The theory of quantum computation offers a number of strategies
for preserving coherence of quantum evolution in the presence of a
decohering environment. To be sure, the idea of classical error correction 
\index{classical error correction} of simply
storing information redundantly and measuring the full system to decide 
with a majority rule whether an error has occurred does not work; 
in that case the measurement itself necessarily destroys the coherence in the system,
as it acquires information about the encoded quantum state.  
It was one of the major breakthroughs in the field of quantum computation 
that quantum error correction could nevertheless be realized. 
One indeed encodes quantum information in several physical systems, but
in a way that  in later partial measurements, one can only 
infer whether an error 
has occurred or not, 
but without being  able to gather 
any information about the encoded state itself \cite{Shor,Steane}.
Based on this knowledge, the error can then be corrected. 

The idea of quantum error correction has been further developed into the 
theory of fault-tolerance \cite{Aharonov,Fault}: \index{fault tolerance}
Even using faulty devices, an arbitrarily long quantum computation can be executed
reliably. In topological quantum memories, \index{topological quantum memories}
systems are arranged in a two-dimensional array on a surface of 
nontrivial topology \cite{Topo}. In physical systems, all these
ideas may further be enhanced with
ideas of trying to stay within physical decoherence-free subspaces \cite{DFS}, 
\index{decoherence-free subspaces}
or bang-bang control. In the debate, Stuart Hameroff said:
\begin{quotation}
I mentioned [yesterday] that micotubules seem to have used the Fibonacci series in 
terms of their helical winding and it has been suggested that they utilize topological 
quantum error correction codes that could be emulated in [man-made] technology. 
As far as redundancy there's a lot of parallelism in the brain and memory seems to be 
representable holographically, so redundancy is not a problem.
\end{quotation}
So why should, after all, 
nature not operate the brain as a fault tolerant quantum computer?

Although this is a tempting idea it is by far more
appealing that it is a realistic option. Beautiful as the idea is, it only works
if the basic operations (called gates) \index{quantum gates} are not too faulty. In realistic terms, 
they have to be 
very, very good. Specifically, quantum fault tolerance, employing complicated 
concatenated encoding schemes \cite{Aharonov,Fault}, works if the performance of logic operations is better than a certain finite 
threshold. If the probability of failure of 
a basic logic operation is below this threshold, then  a computation can indeed
be performed as if perfect quantum gates were available. To 
obtain good upper and lower bounds to 
the exact value of this threshold is a topic of intense 
research, but values of about $10^{-3}$ are realistic. Presently, 
we are a long way from achieving 
such low probability of error experimentally, even in 
sophisticated systems of laser cooled ions in traps, or in 
optical systems. To say, as Hameroff did in the public debate, that 
\begin{quotation}
[...] if you add the potential effect of topological quantum error correction
you get  an indefinite extension,
\end{quotation}
 misses the point that such quantum error
correction is only possible once you have already reached the regime of very small errors.
The required accuracy is 
in very sharp contrast to any accuracy that seems plausibly to 
be available in the slightly above room temperature environment 
of the brain. To  think of performing reliable arbitrarily long quantum 
computation under these conditions is frankly unrealistic. Thus while 
the appeal of fault tolerance as an argument in favor
of large scale coherence is indeed enormous, the numbers very 
strongly argue against that.

\subsection{Uselessness of quantum algorithms for organisms} \index{quantum algorithms}

A final objection to the idea that quantum computing in the brain would have evolved through natural selection is that it would not be useful. Quantum computing has no advantage over classical computing unless it is done on a large scale \cite{NieChu00}. It is difficult to make statements about the time scales for quantum operations in the brain because there is zero evidence for their existence, and because existing platforms on which quantum computing is being explored are immensely different from any known biological system. But for no other reason than the difficulty in doing quantum error correction compared to classical error correction, \index{classical error correction} it can only be expected that the time required to do a quantum logic operation would be greater than the corresponding time for classical logic operations. Because of this, quantum computing to solve any given problem would actually be slower than classical computing until the problem reaches some threshold size.  

History is littered with case studies of organs and attributes that seem to defy Darwinian evolution because any intermediate stage on the path towards their full development would be useless or even harmful. But none have stood up to scrutiny \cite{Daw04}. So perhaps the hypothetical quantum computer in the brain could have come into existence despite the above arguments. After all, quantum computers are generally thought to provide an exponential speed up in solving certain problems \cite{NieChu00}, so the threshold problem size needed to overtake the limitations of intrinsically slow quantum logic operations is not so large. Unfortunately, the sort of problems for which such a speed up exists have no obvious application to a biological organism. Basically, the problems quantum computers are really good at are number theoretic in nature. Instances of these problems, such as factoring \index{factoring} large semi-prime numbers, form the basis of modern cryptography as used millions of times a day on the internet (RSA encryption). If it were not for this fact, such problems would be regarded as mathematical curiosities. Do enthusiasts for biological quantum computing imagine that animals evolved the ability to send RSA-encrypted messages to one another, and subsequently evolved the means to eavesdrop by quantum computing? 

To be fair, there are problems of more general use that quantum computers can attack using the 
Grover search algorithm \cite{Grover} \index{search algorithm}
and its relatives \cite{NieChu00}. Grover's algorithm is sometimes 
described as being useful for ``searching a database'',  suggesting that, for example, it would help 
one find a person in an (alphabetically ordered) phonebook if all one had was their phone number.   
This is a misconception.
%; the search algorithm again is useful for solving mathematical problems in which the large ``database'' is generated from some mathematical function that itself must be defined  in terms of a far smaller number of parameters. 
The Grover algorithm is an important quantum algorithm --- indeed it was one of
the breakthrough results --- but it cannot search a classical database. 
What it requires is a quantum database: a fixed, fully hard-wired database-``oracle'', \index{oracle}
a black box that is ``called''  in the process of the quantum algorithm.  
Nevertheless, Grover's algorithm and its relations may be applied to hard problems, such as finding good routes in a network, that 
would conceivably be useful to an animal. Unfortunately, the speed-up offered by Grover's algorithm on such problems is at best quadratic. Moreover, it has been proven that no algorithm can do better than Grover's algorithm. Thus quantum computers make no difference to the complexity class of these problems. The lack of an exponential speed-up means that the threshold problem size for any speed-up at all is very large. This makes it exceedingly unlikely that evolution could have made the leap to large-scale quantum computing.

\section{Quantum computing in genetics} \label{sec:WEgen}

\subsection{Quantum search}

If not in the brain, then perhaps coherent quantum effects, or even fully fledged quantum computations, are operating behind the scenes at the microscopic level of our genes \cite{Dav04a}. 
It has been argued that the genetic code \index{genetic code} 
contains evidence for optimization of  a
quantum search algorithm. \index{search algorithm} 
%Also, nature might have made use of quantum computation 
%to find a ``fast-track to life'', \index{Origin of life} exploring more quickly the huge space of options in possible
%evolutions by means of quantum algorithms. 
Again, this is intriguing idea, and it may not be 
possible at the present stage to definitively rule it out. 
Here we argue, however, that the case for such an idea 
is, if anything, weaker than that for quantum computing in the brain.

The argument formulated, albeit cautiously, in Ref. \cite{Patel}
in favor of quantum effects to play a role in genetics,  is 
to a large extent based on suggestive numbers that are involved:
On the one hand, the genetic code is based on triplets of 
nucleotides of $4$ varieties that code for $20$ or $21$ 
amino acids. On the other hand, the optimal number $Q$
of sampling operations in Grover's algorithm on an unsorted
database of $N$ objects is given by $Q=1$ for $N=4$ and
$Q=3$ for $N=20$ or $N=21$. This might appear indeed
as a remarkable coincidence of numbers. 

But then, some caution is appropriate: To start with,
the role of $Q$ and $N$ is very different. 
More convincing as an argument
against a connection, however,
is probably the observation that $3,4,20,21$ also appear, say,
in the sequence of numbers which appear the same\footnote{To 
represent a given number in base $b$, one proceeds
as follows:
If a digit exceeds $b$, one has to 
subtract $b$ and carry $1$. 
In a fractional base $b/c$, one subtracts $b$ and carries $c$.}
when written in base $5$ and base $10/2$.
This is 
easily revealed by using the On-Line 
Encyclopedia of Integer Sequences of AT$\&$T Research \cite{ATT}.
It is an interesting and educational pastime
to see how essentially every finite sequence
of integer numbers that one can possibly come up with
appears in, for example, the ``number of isolated-pentagon 
fullerenes with a certain number of vertices'', or
the ``decimal expansion of Buffon's constant''. The sequence
$2,4,6,9$ in this order, to consider a different random 
example,  appears in no fewer than $165$ (!) listed integer
sequences, each of which is equipped with a different 
construction or operational meaning. The lesson to
learn is that one should probably be not too surprised about
coincidences of small tuples of integers.

Moreover, as has been emphasized above, Grover's search is not an
algorithm that sorts a database given as a number of objects following
the laws of classical mechanics: One needs a hard-wired 
oracle, following the rules of quantum mechanics between all involved objects
throughout the computation \cite{Grover}. 
It is difficult to 
conceive how such a hard-wired coherent oracle would be 
realized at the genome level. 
The optimal improvement in the sampling efficiency, in turn,
would be of the order of the square root of $N$. 
It does seem unlikely that the
overhead needed in a reliable quantum computation, possibly even enhanced by
error correction requiring again an enormous overhead, 
would by any figure of merit be more economical than, say, 
a simple doubling of the waiting time in case of $N=4$.

\subsection{Teleological aspects and the fast-track to life} \index{teleology}

One of the most interesting open questions at the interface of the biological and 
physical sciences is the 
exact mechanism that led to the 
leap from complex molecules to living entities. The path from a non-living 
complex structure to one of the possible living structures 
may in some way be a
regarded as a search procedure, the number of potential 
living structures 
being likely a tiny subset of all possible ones consisting of the same constituents
\cite{Dav04a}. Now, how
has nature found its way to this tiny subset? Needless to say, 
we have very 
little to say about this key question. 
In this subsection, we merely cautiously warn that 
whatever the mechanism, 
the involvement of quantum superpositions 
to ``fast-track'' this search 
again in the sense of a quantum search appears implausible.

When assessing the possibility of quantum search here 
one has to keep in 
mind that quantum search is, once again, not just a 
quantum way of having a 
look in a classical database of options: Necessarily, 
the coherence must be 
preserved.
This means that in the search, the figure of merit, the oracle, needs to be hard-wired.
This oracle has to couple to all subspaces corresponding to all different options of
developments. What is more, there is a teleological issue here: It is not entirely
clear what the search algorithm would be searching for. The figure of merit is not
well defined. If a search is successful, life has been created, but what features
does life have? Arguably, this might be linked to the structure 
being able to reproduce. But again, this figure of merit could only be 
evaluated by considering subsequent generations. Thus it seems that it would be  
necessary to preserve a coherent superposition through multiple generations of 
such structures, which we would argue is particularly implausible. 

\section{Quantum consciousness} \index{consciousness}

\subsection{Computability and free will} \index{free will} \index{computability}

Recent years have seen significant advances in the
understanding of neural correlates of consciousness \cite{Koch}. 
Yet, needless to say, the understanding of consciousness on the
biological elementary level is not sufficiently
advanced to decide the case of 
quantum mechanics playing a role or not in  
consciousness, beyond the obvious involvement of 
ruling the underlying physical laws. Hence, any
discussion on the role of quantum mechanics in form of 
long-range entanglement in the brain or in actual realistic collapses
of wave-functions is necessarily of highly speculative character. \index{wave-function collapse} %We have little
%to add to this speculation, which can indeed be a fruitful avenue to follow to 
%explore the realm of possibilities.  
Here, we limit ourselves to addressing arguments 
put forward in the 
public debate which triggered the publication of this book, 
and warn of the possibility of fallacies in some of these arguments.

Where could quantum mechanics play a key role in consciousness?
Hameroff \index{Hameroff, Stuart} argued in the debate, based on an earlier proposal put forth in Ref.\
\cite{Ham}, that the gravitational induced collapse of the wave-function is eventually
responsible for conscious acts. Moreover, microtubules \index{microtubules} forming the cytoskeleton
of neurons should be the right place to look for such state reductions. These reductions should be
realistic, actually happening state reductions, in what 
is called an orchestrated objective reduction (Orch-OR).
\index{orchestrated objective reduction}

This is interesting, but also dangerous territory.
To start with, it does not refer to the established physical theory of quantum 
mechanics as such \cite{Grush,Penrose}.
 The motivation for this approach is to seek a way for human consciousness
to be noncomputable, in order to differentiate it from mere 
computation as performed by aritifical intelligence machines 
(see also Sec.~\ref{freewilly}).
 But computability and noncomputability are the same in quantum 
computer science as in classical computer science. Thus Penrose \index{Penrose, Roger} and Hameroff 
must appeal to a new theory of nature that may allow for noncomputible physical 
effects. They speculate that the key feature of this 
new theory would result from unifying quantum mechancis with general relativity (i.e.
gravity).  \index{general relativity} \index{quantum gravity}

There is no generally accepted theory of 
quantum gravity. Hence, to invoke a realistic collapse in this sense 
bears the risk that the debate is pushed into a dark corner where 
everybody simply has to admit that he or she has no proper 
understanding what is happening there: Ha, I told 
you that you do not 
know the answer! In the debate, Hameroff invoked the  
\begin{quotation}
[...] hypothesis of quantum gravity, which is the only way out of us being helpless
spectators,
\end{quotation}
(that is, the only way to prevent our thoughts from being computible). The mere wish that gravity could leave a loophole for free will  \index{free willy}
does not seem to us to be a very strong argument for this hypothesis.  
Finally, it should be pointed out that there is essentially 
no experimental evidence for any sort of information processing in 
microtubules, still less quantum information processing, and yet less again 
for noncomputible quantum gravitational information processing. 

\subsection{Time scales} 

Invoking quantum gravity also leads to confusions in the  
comparison of time
scales relevant for coherent quantum effects. In the debate, 
Hameroff said: \index{Hameroff, Stuart}
\begin{quotation}
One of these guys [on the affirmative team] 
mentioned that how seemingly ludicruous it is to bring in
quantum gravity because it is $24$
orders of magnitude lower than decoherence\footnote{For
an actual comparison of the relevant time scales, 
see Ref.\ \cite{Hagan}.}. 
\index{decoherence time}
However, in Roger's scheme the reduction is instantaneous so the power is
actually calculated as a kilowatt per tubulin protein.
\end{quotation}
To this Zeilinger (also on the negative team) 
asked \index{Zeilinger, Anton}
\begin{quotation}
But why don't we all boil if it is a kilowatt?
\end{quotation}
to which the response was
\begin{quotation}
Because it is only over a Planck time $10^{-42}$ seconds.
\end{quotation}
These statements refer to the postulated 
Orch-OR time scale of state vector reduction. The relevant decoherence
time scales are given in Ref.\ \cite{Hagan}; this collection 
of numbers contains on the one hand estimates for 
environment-induced decoherence times, for example of a 
superposition of neural firing ($10^{-20}$ seconds).
On the other hand, it gives the time scale of superposition decay in Orch-OR, 
$10^{-4}$--$10^{-5}$ seconds. Based on these numbers, 
the obvious conclusion would be that, since the gravitationally induced 
Orch-OR time scale is so much slower 
than decoherence, the former process will be basically irrelevant.

What is more, the status of these two numbers is very different:
The environment-induced decoherence time scale is
calculated with the help of
standard quantum mechanics as could be tought in any
second year quantum mechanics course. In contrast, the 
number on  Orch-OR  derives from a 
speculative reading of what effects quantum gravity could 
possibly play here. In this figure in Ref.\ \cite{Hagan}, 
these two numbers are put together on the same footing, 
written in the same font size. There is nothing wrong with openly speculating, and the 
presented approach is 
not necessarily wrong or uninteresting. But it can become problematic when the 
right disclaimers are not put in the right places, where
speculation on time scales of a potential theory of gravity are discussed with
the same words and on the same footing as an elementary 
standard quantum mechanics calculation. Regarding 
the status of the $10^{-4}$--$10^{-5}$ seconds
it is not even entirely clear what object it refers to. Also, 
the fact that the conscious thinking process occurs on similar time 
scales to this hypothetical 
Orch-OR, does not 
make the processes causally linked. To make that link is to risk introducing  a rather 
postmodern \index{postmodernism} tone into the debate, where ``anything goes''.

\section{Quantum free will} \index{free will} \label{freewilly}

\subsection{Predictability and free will} \index{predictability}

As mooted in the introduction, there is a relation between life and quantum physics that may motivate a 
philosopher, \index{philosophy} if not a biologist, to try to understand advanced quantum physics.
This is the fact that quantum physics implies an in-principle distinction between (classical) digital computers \index{digital computation} and human brains: the behavior of the former is predictable, while that of the latter is not. Note that we are not just making the obvious observation that in practice the actions of human beings are unpredictable; we are making the stronger statement that no matter how well you observed your neighbor 
(and your neighbor's surroundings), with the help of any level of technology, 
and how well you understood them, with the help of any level of computing power (including quantum computers!), \index{quantum computation} you could not predict precisely how they would respond to a given stimulus (such as your kicking a ball into their yard) at some point in the sufficiently distant future.

Digital computers are designed to have deterministic outputs for a given input. Apart from hardware errors, which happen very infrequently, the output of computer is completely predictable simply by feeding the input into an identically designed machine. Human brains are not designed at all, but more to the point they are analog devices. \index{analog computation} Moreover, they are extremely complicated systems, comprising roughly $10^{11}$ neurons, each electrically connected with varying strength to many other neurons. \index{neuron} And each neuron is a non-trivial element in itself, with complex biochemical reactions determining how it responds to its stimuli. Thus there is every reason to expect the brain to be a chaotic system, \index{chaotic dynamics} in that a small difference to the initial microscopic conditions of the brain would be amplified over time so as to lead to macroscopically different behavior (such as kicking the ball back, or throwing it back).

The above argument does not yet establish an in-principle difference between brain and computer, because in principle it would seem that a sufficiently advanced technology would allow you to know the microscopic state of your neighbor's 
brain (and the microscopic state of their body and other surroundings) to any degree of accuracy, 
so that in principle its state at some fixed future time could be be predicted to any degree of accuracy. What prevents this is of course quantum mechanics: it is impossible to know precisely the position and momentum of a particles. Under chaotic dynamics, this microscopic quantum \index{quantum uncertainty} uncertainty will be amplified up to macroscopic uncertainty. Even for a huge system with few degrees of freedom --- Saturn's moon Hyperion \index{Hyperion} --- the time taken for its orientation to become completely unpredictable according to quantum mechanics 
is only 20 years \cite{Zur98}. For a far smaller and far more complex system such as the human brain, we would expect this time to be far, far smaller --- see also Ref.~\cite{Den84}. 

Thus quantum mechanics implies that, even if artificial intelligence \index{artificial intelligence} were realized on a classical digital computer, it would remain different from human intelligence in being predictable. Of course this does not mean artificial intelligence would be deficient in any aspect of human intelligence that we value, such as empathy or the ability to write poetry. However, such an artificial intelligence would lack free will, at least in the following 
operational sense: If it thought that it had free will, then it would make the wrong decision in Newcomb's problem \index{Newcomb's problem}, by thinking that it could outwit a Predictor of its behavior \cite{Noz69}. For humans, by contrast, such a Predictor cannot exist, except as a supernatural being (a case we have no call to address). 

\subsection{Determinism and free will} \index{determinism}

Having made this distinction between human brains and deterministic digital computers, it is important to note that the above arguments do {\em not} mean that human brains are non-deterministic (still less that they are uncomputable, as Penrose feels they must be \cite{Pen90}). \index{computability} \index{Penrose, Roger} The reason is that determinism and in-principle predictability are not the same things. There are deterministic theories in which systems are unpredictable even in principle because there are in-principle limitations on how much any physical observer can find out about the initial conditions of the system. Moreover, these theories are not mere philosopher's toys. 
\index{philosophy} One of the more popular interpretations of quantum mechanics, known as Bohmian mechanics \cite{Boh52,BohHil93,Hol93,CusFinGol96}, is just such a theory.\footnote{Please note that the following discussion reflects only the opinions of one of us (HMW).} \index{Bohmian mechanics}

In the Bohmian interpretation of quantum mechanics, \index{interpretation of quantum mechanics} quantum particles have definite positions which 
move guided by the universal wave-function $\Psi$. The latter evolves according to \sch's equation; it never collapses. All that ``collapses'' \index{wave-function collapse} is an observer's knowledge of the positions of particles, and this ``collapse'' is nothing but Bayesian \index{Bayesianism} updating based on correlations between the particles in the system of interest and the particles from which the observer is constituted (and on which the oberserver's consciousness supervenes). \index{consciousness} Because of the way the particles' motion is guided by $\Psi$, it can be shown that the observer's knowledge of the position $x$ of a particle for some system is limited by quantum uncertainty \index{quantum uncertainty} in exactly the same way as 
%can never be more precise than that expressed in the probability distribution $|\psi(x)|^2$. (Here %$\psi(x)$ is an {\em effective} wave-function that the observer assigns to that system, just as in 
in orthodox quantum mechanics. \index{orthodox quantum mechanics} But, since Bohmian mechanics is a deterministic theory, probability enters only through observer's lack of knowledge about the position of particles,
due in part to their chaotic motion \cite{Val91}.\index{chaotic dynamics}

In the biological context, this interpretation says that the behavior of humans is determined, by the initial positions of the particles in the person's brain, and its environment. The latter is naturally regarded as a random influence, while the former is more naturally regarded as the source of an individual's will. It is impossible for an outside observer, no matter how skilled, to find out precisely the positions of the particles in an individual's brain, without making a precise quantum measurement of the positions. Such a measurement would instantly destroy the brain by creating states with unbounded energy. Thus, in the Bohmian interpretation, the actions of an individual derive from the physical configuration of their brain, but quantum mechanics makes this configuration unknowable  in-principle to anyone else. For compatibilists \index{compatabilism} \footnote{That is, those who hold that determinism is compatible with --- or even a precondition of --- free will \cite{Den84}.}, the 
picture offered by Bohmian mechanics --- a deterministic yet 
unpredictable {\em quantum free will} --- may be an appealing one. 

%This is the one thing where philosophically minded biologists may be interested, and could be argued to be non-trivial. There is an in-principle difference between computer consciousness and biological consciousness. Digital computers are designed to have deterministic outputs for a given input. Biogical systems will be choatic. Doesn't that just mean we need to know the initial conditions better? No, because of QM. Arguments a la Zurek. Expect the time to be much shorter than for hyperion because its an enormously complicated system. So in principle, no physical observer can predict your behaviour. Not rule out determinism however: e.g. Bohmian mechanics - this could be regarded as theory of the soul, if the soul is something unknowable  that determines our actions. Recall question asked to Peres (I think): in Q teleportation, is the soul teleported, or only the body? Peres said: only the soul, not the body. This is cute, but in the above theory of the soul, the answer would be (as the reporter expected), only the body, not the soul.  

\section*{Acknowledgements}
HMW acknowledges discussion with Eric Calvicanti \index{Calvicanti, Eric} regarding free will in quantum mechanics. 
HMW was supported by the Australian Research Council Federation Fellowship scheme and 
Centre for Quantum Computer Technology, and the State of Queensland. JE was supported by the DFG, Microsoft Research, the EPSRC, and the EURYI Award Scheme.

\end{document}